# Ultralow friction of ink-jet printed graphene flakes

R. Buzio[a]*, A. Gerbi[a], S. Uttiya[a,b], C. Bernini[a], A. E. Del Rio Castillo[c], F. Palazon[d], A. S. Siri[a,b], V. Pellegrini[c], L. Pellegrino[a], and F. Bonaccorso[c]*

[a] CNR-SPIN Institute for Superconductors, Innovative Materials and Devices, C.so Perrone 24, I-16152 Genova, Italy.

[b] Physics Department, University of Genova, Via Dodecaneso 33, I-16146 Genova, Italy.

[c] Istituto Italiano di Tecnologia, Graphene Labs, Via Morego 30, I-16163 Genova, Italy.

[d] Istituto Italiano di Tecnologia, Nanochemistry Department, Via Morego 30, I-16163 Genova, Italy.

*Correstponding authors: renato.buzio@spin.cnr.it; francesco.bonaccorso@iit.it

## Abstract

We report the frictional response of few-layer graphene (FLG) flakes obtained by liquid phase exfoliation (LPE) of pristine graphite. To this end, we inkjet print FLG on bare and hexamethyldisilazane-terminated $SiO_2$ substrates, producing micrometric patterns with nanoscopic roughness that are investigated by atomic force microscopy. Normal force spectroscopy and atomically resolved morphologies indicate reduced surface contamination by solvents after a vacuum annealing procedure. Notably, the printed FLG flakes show ultralow friction comparable with micromechanically exfoliated graphene flakes. Lubricity is retained on flakes with lateral size of a few tens of nanometres, and with thickness as small as ~ 2 nm, confirming the high crystalline quality and low defects density in the FLG basal plane. Surface exposed step edges exhibit the highest friction values, representing preferential sites for originating secondary dissipative processes related to edge straining, wear or lateral displacement of the flakes. Our work demonstrates that LPE enables fundamental studies on graphene friction to the single-flake level. The capability to deliver ultralow-friction-graphene over technologically relevant substrates, using a scalable production route and a high-throughput, large-area printing technique, may also open up new opportunities in the lubrication of micro- and nano-electromechanical systems.

# Introduction

The ability to isolate single atomic layers and few-layer flakes of layered solid lubricants, such as graphite, molybdenum disulfide ($MoS_2$) and hexagonal boron nitride (*h*-BN), has promoted active research into the frictional properties of two-dimensional (2D) materials.[1] With respect to their bulk counterpart, 2D materials offer the advantage of easy deposition and conformation onto surfaces due to their planar structure. In this context graphene has drawn considerable attention, because of its chemical inertness,[2] mechanical stiffness (1060 GPa),[3,4] large out-of-plane flexibility[5] and easy shear capability,[1] which are considered major attributes to achieve low friction[6–8] and low wear rates.[6,8–10] To date, the majority of the fundamental studies on graphene friction have been carried out by atomic force microscopy (AFM), mainly using micromechanically cleaved graphene[11] in form of large micrometric flakes.[12–21] Alternatively, graphene grown either by sublimation of SiC[22,23] or by chemical vapour deposition (CVD)[24] has been tested.[25–28] It has been shown that nanoscale friction forces are a sensitive gauge of both the graphene surface corrugation and the graphene-substrate interface nature.[1] When single layer graphene (SLG) strongly adheres to an atomically-smooth substrate (*e.g.*, *h*-BN, mica),[13] the AFM tip experiences ultralow friction forces comparable with those recorded on bulk graphite (roughly from ~0.1 to ~1nN friction forces for normal loads up to several tens nN, depending on system details).[12,18] Weak adhesion due to interfacial atomic roughness causes out-of-plane puckering of the supported SLG around the tip, thus leading to larger contact areas and enhanced friction by factors ~2-3.[12,13] On such weakly adherent substrates (*e.g.*, $SiO_2$), the lowest friction values are only measured on the thicker multilayers graphene (MLG) (*e.g.*, eight-layers or above),[13] which are less susceptible, with respect to SLG, to out-of-plane deformation.[12,13] Besides substrate effects, the ultralow friction property of graphene can be dramatically deteriorated by chemical functionalization (*e.g.*, with $O_2$, F, H),[15,17,20,26,28] surface defects,[29] roughness (*e.g.*, wrinkles, exposed step edges),[14,18,19] and contamination (related to graphene production and deposition processes).[21] In fact, all these factors increase the corrugation of the tip-graphene interfacial potential and impact the graphene out-of-plane bending stiffness, enhancing energy dissipation by factors ~2-10[1,15,28] or even higher.[30]

Recently, a number of reports focused on the friction and wear properties of solution-processed graphene,[31] dispensed to macroscale rotating/sliding contacts[10,32–37] *via* liquid dispersions of FLG flakes. Compared to cleaved or bottom-up grown graphene,[31] dispersions of FLG flakes are in general more suitable for practical applications, such as printed and flexible electronics or where large scale production is a must, such as in the composite, tribology and energy fields.[10,32–37] Moreover, dispersions offer a new platform for fundamental studies on graphene lubricity.[37] Amongst the diverse methods exploited to produce and process SLG and FLG flakes in liquid environment,[31] it is crucial to identify those offering balance between the ease of fabrication/deposition and the flakes quality. This selection process is greatly assisted by AFM,[38] which can probe different sources of solution-processed FLG flakes against the well-established benchmark of SLG/FLG produced by micromechanical cleavage (MC).[12–21] Liquid phase exfoliation (LPE) is an economic, easily scalable alternative to MC[11] and bottom-up approaches,[22–24] which must be considered when mass production and cost reduction are key parameters.[31,39] The LPE process exploits the dispersion of graphite crystals in a solvent, and by applying energy to the mixture (sonic waves,[40–44] mechanical milling,[45] shear mixing[46]) produces dispersions composed by a mixture of SLG/FLG as well as MLG flakes. The solvent used can be water or an organic solvent,[47,48] with (or without) exfoliation agents such as polymers,[49] surfactants[41,42,50] or ionic liquids.[51,52] These exfoliating agents prevent the flakes re-stacking due to the strong π-π interaction.[53] The use of the exfoliating agents has shown to be ideal to obtain high concentration (> 5gL$^{-1}$) of dispersed flakes.[48] The presence of surfactants, polymers or ionic liquids can be a problem if the final application requires a "pure" material, as is the case of electronic applications.[54,55] The LPE of graphite performed in organic solvents avoids this problem, because the solvent can be removed by evaporation.[40,55] The selection of the solvent is crucial for the exfoliation process.[31,39] In fact, it must have the "right" surface tension (46.7 mNm$^{-1}$)[56] equivalent to the surface energy of graphene. The LPE process can produce SLG and FLG flakes with crystalline quality and lateral size varying from a few tens[57] up to several hundred of nanometers.[41] Practical issues however make the surface properties of such flakes much more subtle and complex compared to MC and synthetically grown SLGs and FLGs.[58–61] Also, trace amounts of high-boiling-point solvents can remain trapped between the flakes, and may contaminate their surfaces.[40,62] Given the sensitivity of interfacial friction to fine details of the graphene structure and graphene-substrate interaction, the appearance of a ultralow friction state in contact junctions lubricated by LPE graphene is far from being obvious. A quantitative AFM study of state-of-the-art graphene-based inks provides a timely and direct strategy to address the aforementioned issue.

In this work, we probe by AFM the frictional response of FLG flakes obtained by LPE of graphite in conventional organic solvents and dispersed *via* inkjet printing[55,63] onto both $SiO_2$ and hexamethyldisilazane (HMDS)-terminated $SiO_2$ substrates. We systematically contrast ink-jet printed and MC SLG/FLG flakes *via* load- and thickness-dependent friction measurements. The friction response is explored for inks exploiting two different solvent-graphene interactions, namely with N-methyl-pyrrolidone (NMP)[40] and ortho-dichlorobenzene (ODCB).[62,64] The main steps of the experimental methodology are summarized in Fig. 1.

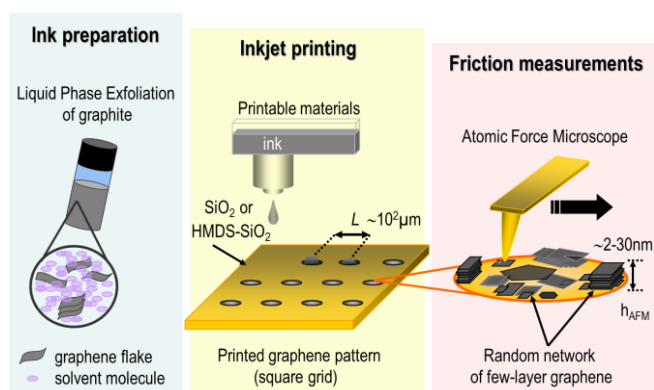

**Fig. 1.** Schematic illustration, from left to right, of the ink preparation, ink-jet printing process and friction measurements by AFM, respectively.

# Experimental

## Materials and liquid phase exfoliation procedure

The graphite flakes (+100 mesh, ≥75 % min), NMP (99.5 % purity) and ODCB (99.5 % purity) are purchased by Sigma-Aldrich and used without further purification.

We exploited LPE of graphite to produce the graphene inks in NMP and ODCB.[55] Experimentally, two batches of 1 g of graphite flakes are dispersed in 100 mL of NMP and ODCB, the batches are ultrasonicated (in a VWR ultrasonic bath) for 6 hours. The resulting dispersion is then ultra-centrifuged at ~16000 $g$ in a Beckman Coulter Optima™ XE-90 (SW32Ti rotor), and by exploiting sedimentation-based separation[55,57,64] we removed the thicker flakes and un-exfoliated graphite flakes. Subsequently, we collected 90% of the supernatant by pipetting. The ink-jet printing in general requires concentrated (> 0.3 gL$^{-1}$) inks to fit the printing requirement of viscosity,[55,63,65–67] and to minimize the number of printing passes.[55] To achieve this target concentration, the supernatant extracted is ultra-centrifuged at ~ 200000 $g$. The high $g$ force promotes the sedimentation of the SLG/FLG/MLG flakes taking advantage of the higher density of the graphitic flakes (~2.1g/cm$^3$)[68] with respect to the solvents densities ($\rho_{NMP}$ = 1.03 g/cm$^3$, $\rho_{ODCB}$ = 1.3 g/cm$^3$).[69] The sedimented graphitic flakes are collected and re-suspended in 3mL of their respective solvents, *i.e.*, NMP and ODCB, by using an ultrasonic bath for 10 min.

## Characterization of the graphene inks

**Optical Absorption Spectroscopy.** Optical absorption spectroscopy (OAS) of the as-produced inks is carried out in the 300-1200 nm range with a Cary Varian 5000i UV-*vis*-NIR spectrometer. The absorption spectra are acquired using a 1 mL quartz glass cuvette. The ink is diluted to 1:100 in NMP (or ODCB according to the solvent of the as-produced sample), to avoid scattering losses at higher concentrations.[55] The corresponding solvent baseline is then subtracted. The concentration of graphitic flakes is determined from the optical absorption coefficient at 660 nm, using $A = \alpha l c$ where $l$ [m] is the light path length, $c$ [gL$^{−1}$] is the concentration of dispersed graphitic material, and $\alpha$ [Lg$^{−1}$m$^{−1}$] is the absorption coefficient, with $\alpha$ ~1390 Lg$^{−1}$m$^{−1}$ at 660 nm.[41]

**Rheological measurements.** The viscosity of the inks is measured with a Discovery HR-2 Hybrid Rheometer (TA instruments), using a double-wall concentric cylinders geometry (inner diameter of 32 mm and outer diameter of 35 mm), designed for low-viscosity fluids. The temperatures of the inks are set and maintained at 25º C throughout all the measurements.

**Transmission electron microscopy (TEM).** The morphology of the flakes dispersed in the inks is characterized by using a TEM JOEL JEM 1011, with an acceleration voltage of 100 kV. For the sample preparation, the NMP inks are diluted (1:100) in fresh NMP and 20 μL are drop on copper grids (200 mesh), and dried in vacuum overnight. Lateral size measurements are carried out using Image-J software on 100 flakes per each sample.

**Ink-jet printing of graphene-based inks.** Graphene-based inks, both in NMP and ODCB, are printed onto Si(100) wafers (10×5×0.5 mm$^3$, *n*-type, 1-10 Ohm·cm) by Crystec GmbH, terminated with 300 nm-thick SiO$_2$. The wafers are preliminarily cleaned by bath sonication in acetone and ethanol and wet etching (for 30 s) in buffered hydrofluoric acid by Sigma-Aldrich. Surface modification with HMDS (Sigma-Aldrich) is accomplished by exposing the cleaned substrates to the vapours of ~1 ml HMDS in an evacuated vacuum chamber at 80° C.

Ink-jet printing is carried out on bare $SiO_2$ and HMDS-$SiO_2$ at 30° C using a Fujifilm Dimatix Materials Printer (DMP 2831) with a cartridge designed for a 10 pL nominal volume. Graphene-based flakes are printed according to a simple pattern represented by a square grid of dots (Fig. 1). The grid extends over millimetre squared areas and has surface roughness in the nanometres range. The as-printed patterns are thermally post-annealed at 350°C in high vacuum (< $2\times10^{-6}$ torr, 90 minutes) in order to diminish the presence of solvent residues left after the printing process.[40,62,70]

**Drop-cast deposition of graphene inks.** As-prepared graphene-based inks are drop-cast onto Si(100) wafers ($10\times5\times0.5$ mm$^3$, *n*-type, 1-10Ohm·cm) by Crystec GmbH, terminated with 300nm-thick $SiO_2$. Sedimentation for 24 hours is followed by drying with $N_2$ and vacuum annealing (see section 2.3.1 for annealing details).

**Micromechanically cleaved graphene.** Micromechanical cleavage is carried out as explained elsewhere.[11,13,38] Briefly, highly oriented pyrolytic graphite by NT-MDT (ZYB quality) is exfoliated under ambient conditions *via* Scotch adhesive tape and flakes are transferred to Si(100) substrates terminated with 300nm-thick thermal oxide $SiO_2$.

**Raman spectroscopy.** The as-prepared graphene-based inks are drop-cast onto Si wafers with 300 nm of thermally grown $SiO_2$ (LDB Technologies Ltd.) and dried under vacuum. Raman measurements are collected with a Renishaw inVia confocal Raman microscope using an excitation line of 532nm (2.33eV) with a 50× objective lens, and an incident power of ~1 mW on the sample. 20 spectra are collected for each sample. Peaks are fitted with Lorentzian functions.

**X-ray photoelectron spectroscopy.** Samples for X-ray photoelectron spectroscopy (XPS) measurements are prepared by drop casting the ink onto thermally grown $SiO_2$/Si wafers and measured with a Kratos Axis Ultra DLD spectrometer, using a monochromatic Al Ka source (15 kV, 20 mA). Wide scans are acquired at analyser pass energy of 160eV. High-resolution narrow scans are performed at constant pass energy of 10 eV and steps of 0.1 eV. The photoelectrons are detected at a take-off angle ⍺ = 0° with respect to the surface normal. The pressure in the analysis chamber is maintained below $6.75\times10^{-9}$ Torr for data acquisition. The data are converted to VAMAS format and processed using CasaXPS software, version 2.3.16. Fittings of the spectra are performed using a linear background and Voigt profiles.

**Atomic force microscopy.** Nanoscale surface imaging and force-spectroscopy are carried out under ambient conditions with a commercially available AFM (Solver P47-PRO) system by NT-MDT. Samples are scanned with rectangular shaped Si cantilevers with nominal stiffness 0.3 Nm$^{-1}$ (MikroMasch, CSC37/Al:BS). Calibration of normal and lateral forces is carried out according to the methods proposed by Sader *et al.*[71] and Varenberg *et al.*[72] Force *vs* displacement curves are obtained by recording the cantilever deflection, *i.e.*, the applied normal force $F_N$, while ramping the relative distance between tip and sample and they are transformed into force vs distance ($F_N$ vs $D$) curves by assigning $D=0$ to the region where normal force becomes strongly repulsive.[73]

Lateral force maps are typically acquired at ~1 Hz scan rate and are used to obtain friction force vs normal load ($F_{fric}$ vs $F_N$) characteristics. To this end, $F_N$ is decreased every ten lines from ~35 nN to the pull-off value; the corresponding lateral forces, resulting from the difference between forward and backward scans, are averaged on eight lines in between the $F_N$ jump to produce one data point.[73–75]

The nonlinearity of experimental $F_{fric}$ vs $F_N$ curves is traced back to a nonlinear dependence of the tip-sample contact area $A$ on $F_N$, $A \equiv A(F_N)$, thereby assuming a constant interfacial shear strength $\tau$.[76] In particular the $A \equiv A(F_N)$ relationship is predicted by the continuum Maugis-Dugdale (MD) transition model in its generalized form.[77] This model determines the position of the interface along a spectrum of contact behaviours ranging between two limiting cases, hard contacts with long-range attractive forces or soft contacts with short-ranged adhesion. The former are usually modelled through the simplified Derjaguin-Mueller-Toporov (DMT) theory,[75,78] whereas the latter via the Johnson-Kendall-Roberts (JKR) model.[79,80] In turn, the MD transition model comprehensively treats both limiting and intermediate cases. The location of the contact is embodied by the dimensionless parameter λ which varies nonlinearly from zero to infinity.[77] In practice, for λ<0.1 DMT model applies whereas for λ > 5 the JKR model is appropriate. Intermediate cases have 0.1< λ < 5. To obtain estimates of the transition parameter λ, interfacial shear strength $\tau$ and adhesion energy $\gamma$ at the AFM tip-sample junction, we fit $F_{fric}$ vs $F_N$ with the analytical solution of the transition model developed by Carpick et al.[76] The Young's modulus and Poisson's ratio used in the fit are respectively 70 GPa and 0.2 for the $SiO_2$ substrate, 130 GPa and 0.27 for the Si tips, 30 GPa and 0.24 for the multilayer flakes.[16] Tip apexes, monitored post-experiment by scanning electron microscopy, are of about 30 nm. This value is larger than the nominal radius of a new probe (~10 nm) and indicates tip wear in the course of the experiments. Wear likely takes place during the prolonged scanning required to locate FLG on the high-friction substrates. We assume a tip curvature radius $R$ = 20 nm to simultaneously interpolate data on $SiO_2$ and graphene, with an uncertainty of ±10 nm. The use of single-asperity continuum models to analyze AFM friction forces is widespread, and is particularly convenient for comparative purposes. However, they are known to provide an oversimplified picture of the actual interaction between tip and SLG/FLG flakes. We describe these aspects for completeness in the Appendix, section S1.

The corrugation of the tip-surface interfacial potential, $E_0$, is evaluated from friction loops with atomic contrast using the relationship $E_0 \approx a F_{lat}^{max} / \pi$, where $a$ is the stick-slip periodicity and $F_{lat}^{max}$ is the maximum lateral force in the stick portions.[81]

## Results and Discussion

The as-produced inks are characterized rheologically by OAS and viscosity (ν) measurements, and morphologically, by using Raman spectroscopy and TEM. The concentration of the as-prepared NMP-based ink is 0.35 g L$^{-1}$, calculated from the OAS (Fig. 2(a)). The peak at ~266 nm in the absorption spectrum is the typical signature of the van Hove singularity in the graphene density of states[82]. The ν value of the ink at the constant shear rate of 10 s$^{-1}$ is 2.1 mPa s (Fig. 2(b)). For comparison, the ν value of the NMP is 1.59 mPa s. The ν does not vary with increasing shear rate, which is a behaviour typical of Newtonian fluid[83].

The morphology of the flakes is characterized by TEM (see Fig. 2(c) where a representative TEM bright field image is reported). The TEM images allow the estimation of the lateral size of the flakes, with the statistical analysis that shows a main lateral size distribution centred at ~134 nm (Fig. 2(d)).

The Raman spectrum of the as-prepared ink (Fig. 2(e)) as well as the spectrum of the starting material, *i.e.*, graphite, is shown for comparison. The typical Raman spectrum of the graphene flakes consists in the G peak which is originated from the $E_{2g}$ phonon at the Brillouin zone centre[84]. The D peak corresponds to the breathing modes of $sp^2$ carbon rings, requiring a symmetry breaking for its activation.[85] The 2D peak is the second order of the D peak, appearing as a single peak for SLG and as the superposition of multiple components for FLG and MLG (following the modification of the band structure with the increasing number of layers)[86]. The 2D peak does not require the presence of defects and always appears even in absence of the D mode. A D' peak can also appear, which is a double resonance mode due to an intra-valley process (*i.e.*, connecting two points belonging to the same cone around K or K')[86]. The statistical Raman analysis gives useful information on the quality of the exfoliated graphitic flakes.[55,63] In particular, the relationship of the integral intensity of the D band (normalized with the integral intensity of the G band), I(D)/I(G), plotted against the full width at half maximum of the G band, FWHM(G), gives us information about the type of defects present in the graphene flakes (edges or in-plane defects).[87] The lack of a linear correlation between I(D)/I(G) and FWHM(G), in Fig. 2(f), is attributed to sub-micron lateral sized flakes rather than the presence of structural defects. The 2D peak gives us information about the number of layers, *i.e.*, the 2D peak in bulk graphite consists of two components $2D_1$ and $2D_2$,[86,88,89] roughly the $2D_1$ is half of the $2D_2$ intensity, while for a mechanically exfoliated SLG flake the 2D band is a single and sharp peak, that is roughly 4 times more intense than the G peak.[86] For LPE graphene is uncommon to find such large intensities for 2D, even for SLG.[39–41,52,90] However, an estimation of the thickness of the sample can be made by evaluating the $2D_1/2D_2$ ratio, which in our case indicates that our ink is enriched with SLG/FLG flakes (Fig. 2(g)).

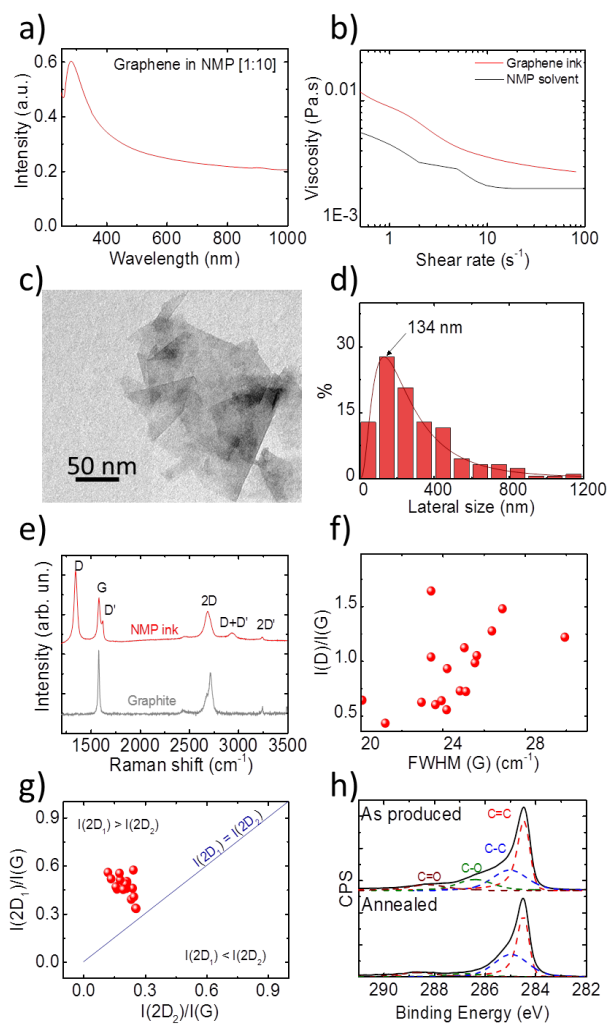

Fig 2. Characterization of the graphene ink in NMP. (a) OAS (diluted 1:10 with NMP), (b) shear rate vs viscosity, (c) representative TEM image of a FLG flake, and (d) lateral size distribution. (e) Raman spectra of FLG flakes (red curve) and graphite (grey curve). Statistical Raman analysis is also shown, in (f) FWHM(G) vs I(D)/I(G), and (g) the ratio I(2D$_2$)/I(G) vs I(2D$_1$)/I(G). (h) XPS analysis of the carbon–oxygen content, showing the effect of vacuum annealing at 350°C on oxygen content reduction.

Additionally, the chemical groups present in the flakes drop casted onto SiO$_2$ are characterized by XPS (Fig. 2(h)). The high-resolution C1s spectrum of the sample, as produced and after annealing at 350° C for 90 min., shows the presence of oxidized C-O and C=O groups at binding energies 286.4 eV and 288.3 eV respectively,[91] probably as a consequence of the exfoliation process. While these groups represent ~29% of the overall carbon content in the as produced sample, this amount is greatly reduced after vacuum annealing. In fact, in the annealed sample, C=O and C-O groups only counts for ~8% of the total carbon content. These groups could also be linked with residual NMP. The evolution of NMP amount on the sample can be estimated from the atomic percentage of nitrogen, whose presence is linked only to the solvent molecules. N 1s spectra (see Appendix, Fig. S7) show that upon annealing, the relative amount of NMP is reduced but not fully removed, with N/(C + N) ratio that goes from ~5.3 at. % to ~3.4 at. %).

The characterization of the ODCB-based ink is reported in the Appendix (sections S6 and S7).

### 3.2 Morphology and friction properties of printed SLG/FLG

The top-view optical micrograph of a typical FLG-based grid, printed on bare SiO$_2$ using an NMP-based ink, is reported in Fig. 3(a). The thickness of the SiO$_2$ creates contrast to make FLG visible under white light illumination.[92] Here, the printing pitch $L$=200 μm is larger than the microdrops spreading diameter, leading to the formation of non-overlapping "coffee ring" deposits of ~100 μm lateral size. Each ring originates from pinning of the contact line of the drying drop and preferential evaporation at the edges.[93] As a consequence, the flakes coverage is not uniform but decreases from the periphery to the centre of the ring. This effect has been previously noted in literature,[63] reflecting the low interaction of graphene with SiO$_2$. Magnification of the centre of the rings reveals micrometric FLG-based patches with dendritic shape, separated by large portions of uncovered substrate (see inset to Fig. 3(a)).

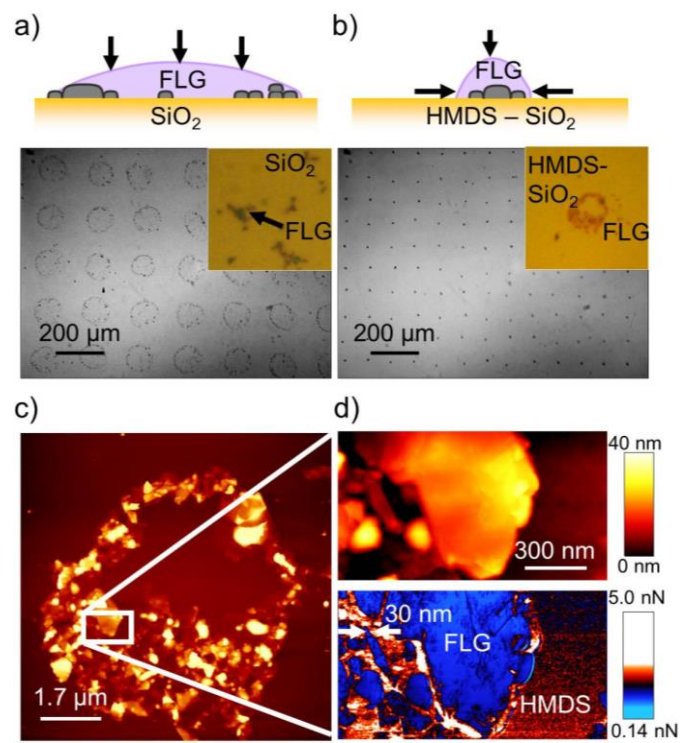

**Fig. 3.** (a) Scheme of a FLG-based ink droplet printed on SiO$_2$. Arrows indicate the vertical shrinking front of the droplet. The grayscale optical micrograph reveals a square pattern of coffee-ring deposits (15 passes). The full color inset (18x18um$^2$) shows the centre of a ring. (b) Scheme for a droplet printed on HMDS-SiO$_2$, showing preferential in-plane shrinking of the ink droplet. The grayscale optical micrograph attests a square grid of micrometric dots (single pass). The inset (20x20um$^2$) shows a single dot. (c) AFM topography of the dot in the inset of (b). (d) Topography and friction maps for the rectangular area in (c) (normal load $F_N$=3nN).

A significant enhancement of the drop in-plane shrinkage takes place when the SiO$_2$ surface energy is reduced through the HMDS functionalization. Fig. 3(b) reveals that in this case the printed grid consists of isolated, micrometric dots of ~10 μm lateral size separated by the printing pitch length, which in this case is $L$=100 μm.

Magnification of the individual dots by optical microscopy (inset of Fig. 3(b)), however, reveals an irregular topology on the micrometric scale, characterized by the coexistence of compact agglomerates and holes that originate from some pinning of the contact line during NMP evaporation. The inhomogeneity of the deposited ink is indeed a common feature already reported in other studies.[20] The random network involves agglomeration and stacking of several individual flakes. As shown in Fig. 3(c) and 3(d), the flakes have their basal plane aligned with the substrate surface and exhibit a lateral extent mostly in the ~30-150 nm range, with one flake above ~300 nm. This size distribution is in line with the TEM data (Fig. 2(c) and Fig. 2(d)). The friction map acquired simultaneously on the same region clearly demonstrates order-of-magnitude reduction of interfacial friction on printed FLG compared to the one registered on HMDS-SiO$_2$ substrate (Fig. 3(d)). Neglecting edge effects (see below), lubricity arises over each individual flake, down to length scales as small as ~30 nm, as shown in Fig. 3(d).

We extensively verified the friction-reduction property of the LPE FLG inks, addressing both printed and drop-casted specimens. Also, we noticed a remarkable friction reduction at the single-flakes level with minimum thickness of ~2.0 nm (see Appendix, Fig. S1 and Fig. 7).

We further explored the surface structure and friction contrast of the printed flakes through atomic-resolution AFM. Representative results are summarized in Fig. 4 for FLG flakes printed onto HMDS-SiO$_2$. Large-scale imaging gives evidence of a layered structure involving at least three flakes stacked together (Fig. 4(a) and Fig. 4(b)).

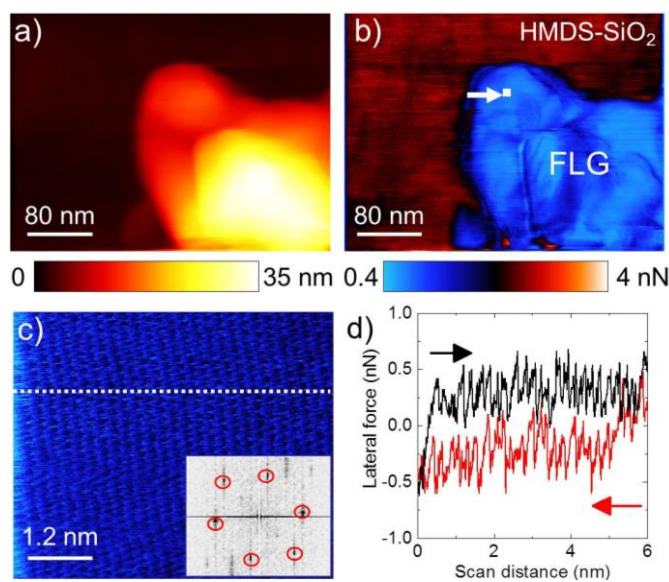

Fig. 4 (a) Topography and (b) friction measured by AFM on a selected FLG aggregate. (c) High-resolution lateral force map on the spot highlighted in (b) by the white arrow (normal load $F_N$=9nN). It shows a threefold symmetric pattern with periodicity ~0.21nm, corresponding to the known graphite lattice. The hexagonal arrangement of peaks in the Fourier transform (inset) gives the local orientation of the graphene lattice with respect to the scan direction. (d) Lateral force along the dashed line in (c) attests a horizontal friction loop with atomic stick-slip motion of the AFM tip.

Examination of the high-resolution lateral force map on a 6×6 nm$^2$ region (Fig. 4(c) and Fig. 4(d)) reveals stick-slip motion with atomic-scale periodicity, similar to that reported for bulk graphite and graphene prepared by MC,[12,13] sublimation of SiC[25] and CVD.[27] The 3-fold symmetric pattern (inset of Fig. 4(c)) corresponds to the known graphite symmetry and gives information on the local orientation of graphene lattice with respect to the scan direction.[27] Inspection of the friction loops (Fig. 4(d)) reveals nearly horizontal force profiles, with stick-slip events of periodicity (~0.21 nm) and amplitude (0.2-0.4 nN) that reasonably agree with studies carried out in similar experimental conditions, *i.e.*, similar tips and loads.[25] The absence of a systematic tilt of the force traces along the scan direction rules out out-of-plane puckering,[12,13] which is expected in view of the large thickness (∼12 nm) of the probed area. We note that friction maps with atomic contrast are often acquired by AFM on printed FLG. This complements XPS data (Fig. 2(h)) and provides an independent evidence that vacuum annealing at 350°C removes most of the solvent residues from the topmost layers of printed FLG.[94]

### 3.3. Spectroscopy of normal and friction forces on printed graphene

We gained deeper knowledge of the friction behavior of printed SLG/FLG onto the bare SiO$_2$ substrate via quantitative spectroscopy of normal and frictional forces. Analysis of force-distance curves attested the formation of elastically-stiff solid-on-solid contacts, similar to those established by the AFM tip with MC graphene[16,95,96] and with the bare SiO$_2$/Si substrates.[73,97] The distribution of the adhesive (pull-off) forces $F_\mathrm{pull-off}$ is however significantly wider for printed FLG. Figure 5 exemplifies the situation for a representative FLG aggregate printed on SiO$_2$ (qualitatively similar results occurred on HMDS-SiO$_2$, see Appendix and Fig. S2). Figures 5(a) and (b) show the morphology and $F_\mathrm{pull-off}$ maps, respectively. It is possible to notice the remarkable difference between the adhesion measured on printed FLG and on SiO$_2$, which originates a distinct contrast for the two surfaces. Examination of the $F_\mathrm{pull-off}$ histogram in Fig. 5(c) reveals that the tip-SiO$_2$ adhesion is approximately Gaussian with mean value ~17 nN and standard deviation ~2 nN, whereas the tip-FLG adhesion displays a wider peak at ~12 nN and a tail of lower values roughly extending from ~2 to 10 nN. The enhanced spread of adhesion on printed graphene, especially below 10 nN, likely reflects force variations from one location to the next induced by topographical roughness.[97] Figure 5(d) shows that force-distance curves have sudden snap-in- and snap-out-of-contact separated by a hard-wall repulsion region.

We extensively inspected normal force spectra at different surface spots without recognizing specific signatures related to a continuous NMP over-layer adsorbed on the topmost FLG flakes. In particular, we did not observe force instabilities after snap-in-contact, associated to the plastic penetration or the squeeze-out of a compliant NMP molecular layer.[73,98] This enforces the conclusion that the trace amounts of NMP detected by XPS (below 4% of nitrogen for vacuum annealed specimens) probably concern molecules localized at the surface exposed step edges, or even subsurface molecules trapped immediately below the topmost layers.[40] Further details on this aspect are reported below (Fig. 7(c) and the related text).

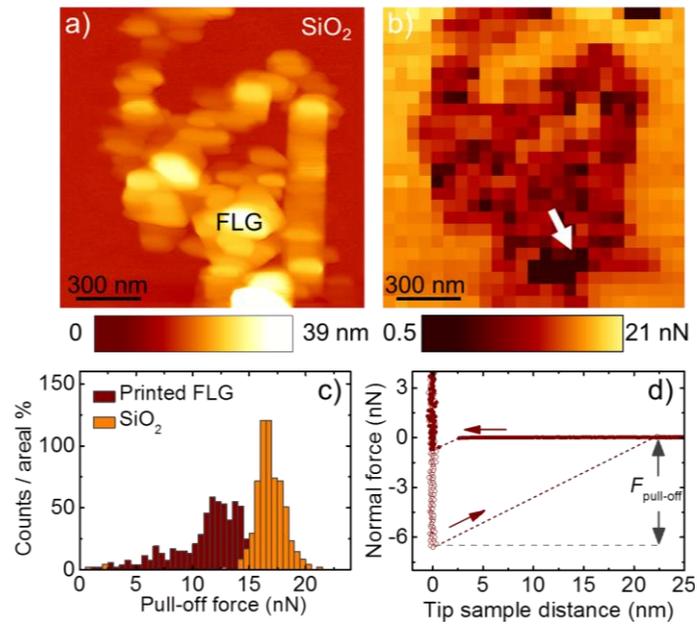

**Fig. 5** (a) Topography of FLG printed on SiO$_2$. (b) Pull-off force map measured on the aggregate in (a), showing distinct adhesion on FLG with respect to the substrate. (c) Histogram of pull-off forces from the map in (b), revealing enhanced spread of adhesion on printed FLG. (d) Force-distance curve acquired at the spot indicated by the white arrow in (b). Force profile indicates the formation of a solid-solid tip-FLG contact.

Variable-load friction measurements were systematically acquired on printed FLG and contrasted with the response of reference surfaces. For convenience, we investigated printed platelet-like aggregates of thickness $h_{AFM} \approx$ 10-30 nm and lateral size above ~200 nm. Results are summarized in Fig. 6(a), and Fig. 6(b) through representative curves acquired on selected surface spots. Figure 6(a) demonstrates that at loads above ~15 nN the friction force on bare SiO$_2$ and HMDS-SiO$_2$ exceeds that on printed FLG by more than 1 order of magnitude. The HMDS treatment clearly lowers interfacial friction of bare SiO$_2$, as it imparts a predominant -CH$_3$ termination to the functionalized surface.[99] However, HMDS also makes the tip-substrate contact mechanically unstable and prone to abrasive wear, that quickly manifests above ~10 nN through slope variations and friction changes after a few scans over the same region (see also Appendix and Fig. S3). This behaviour markedly differs from the friction response of SiO$_2$ and printed FLG, which remained stable and reproducible after several scans over the same region.

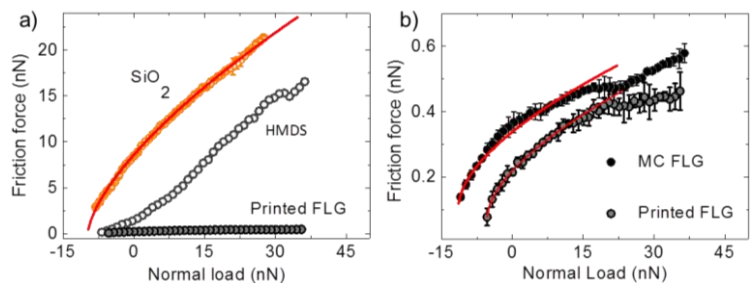

Fig. 6 (a) Experimental friction vs load curves acquired respectively on printed FLG and on the deposition substrates (each curve is an average over 20 individual load-scans). (b) Direct comparison of printed and MC FLG (2 nm-thick) FLG. The red lines are fits with the MD model.

Direct comparison of printed and cleaved FLG flakes is embodied in Fig. 6(b) (see Appendix and Fig. S4 for further details on MC FLG sample). Both surfaces provide comparable friction forces (in the range 0.1-0.6 nN) and friction-load dependence, with the exception of a horizontal shift of ~6 nN indicating a reduction of the pull-off (adhesive) force on printed FLG flakes. Friction interpolation of data on SiO$_2$ (Fig. 6(a)) gives λ~0, hence the tip-SiO$_2$ contact obeys to the simplified DMT theory.[73] With the tip radius $R$=20 nm, one obtains the effective adhesion energy $\gamma_{SiO2} = 78 \text{mJ/m}^2$ and the shear strength $\tau_{SiO2} = 1300 \text{MPa}$. For printed FLG (Fig. 6(b)), fitting procedure yields $\lambda_{FLG,printed} = 1.3$, hence the tip-graphene contact has transitioned towards the JKR regime. Moreover, $\gamma_{FLG,printed} = 55 \text{mJ/m}^2$ and $\tau_{FLG,printed} = 16 \text{MPa}$. Finally, for MC FLG flakes (Fig. 6(b)), $\lambda_{MC FLG} = 1.1$, $\gamma_{MC FLG} = 116 \text{mJ/m}^2$ and $\tau_{MC FLG} \approx 15 \text{MPa}$. The adhesion energy $\gamma_{MC FLG}$ is in the lower end of the available estimates for the SiO$_2$/graphene adhesion energy (96-320 mJ/m$^2$).[16,95,96] More importantly, the relationship $\tau_{FLG,printed} \approx \tau_{MC FLG}$ proves that the LPE FLG flakes show ultralow friction comparable with MC FLG.

Table 1 resumes the outcome of the analysis of experimental curves acquired at several surface spots. The shear strength intervals, $\tau_{FLG,printed} \sim 15\text{-}29 \text{MPa}$ and $\tau_{MC FLG} \sim 12\text{-}24 \text{MPa}$, are largely overlapped. Moreover, they are close to estimates from AFM-based friction experiments and simulations conducted on MC SLG/FLG (14 MPa,[16] 30 MPa[100]) and on highly oriented pyrolytic graphite (10-17 MPa,[74,75] 1-7 MPa[100]). From the τ values of Table 1, the tip-sample effective contact area is easily evaluated as $A \approx F_{fric}/\tau$. Accordingly, $A$ varies between ~2 nm$^2$ and ~13 nm$^2$ on SiO$_2$, when $F_{fric}$ increases from ~2.5 nN to ~20 nN (τ=1520 MPa, Fig. 6(a)). For MC and LPE flakes, $A$ increases from ~5 nm$^2$ to ~25 nm$^2$ when $F_{fric}$ grows from ~0.1 nN to ~0.5 nN (τ≈20 MPa, Fig. 6(b)). Estimates for $A$ are in agreement with the typical nanometric resolution of the AFM morphologies (*e.g.*, see Fig. 3 and Fig. S1).

Table 1. Parameters of the MD transition model obtained by fitting experimental friction vs load curves. The variability range reflects the surface heterogeneity and the ±10nm uncertainty on the tip curvature radius.

| Surface | λ | γ (mJ/m$^2$) | τ (MPa) |
| --- | --- | --- | --- |
| SiO$_2$ | ~0 | 100 ± 50 | 1520 ± 530 |
| Printed FLG | 0.9 ± 0.6 | 71 ± 37 | 22 ± 7 |
| MC FLG | 1.1 ± 0.3 | 160 ± 70 | 18 ± 6 |

We gained additional information on the nature of the FLG flakes-substrate interaction from consideration of thickness-dependent friction curves ($F_{fric}$ *vs* $h_{AFM}$), studied in the limit of few-layers thick samples. It is known that monolayer to 4-layers thick MC flakes supported on SiO$_2$ are subjected to appreciable out-of-plane deformation around the scanning AFM tip, which leads to an increase of the friction force compared to the thicker (≥ 5L thick) flakes.[12,13] This effect is shown in Fig. 7(a) for a MC FLG sample and compared to the response, in Fig. 7(b), of isolated flakes obtained by LPE and deposited by drop-casting onto SiO$_2$ (isolated flakes are in fact rarely observed in ink-jet printed samples). Two facts emerge. First, the minimum apparent height of the LPE flakes, as measured by AFM, is around ~1.3 nm which exceeds by ~0.8 nm the height of MC SLG (~0.5 nm, in agreement with Refs.12,16). Second, the dependence of the friction force on the apparent height is fairly weak for LPE flakes, but not necessarily weaker than that of MC flakes. In fact, since we cannot firmly relate the apparent height to the number of layers for the LPE flakes, as instead done for the MC ones, no straightforward comparison can be drawn on the layer-dependent friction properties for the flakes produced by the two different approaches, *i.e.*, MC and LPE. Nonetheless, we propose that the response of the LPE flakes shown in Fig. 7(b) might be explained by subsurface contamination from residual NMP (see Fig. 7(c) and related text).

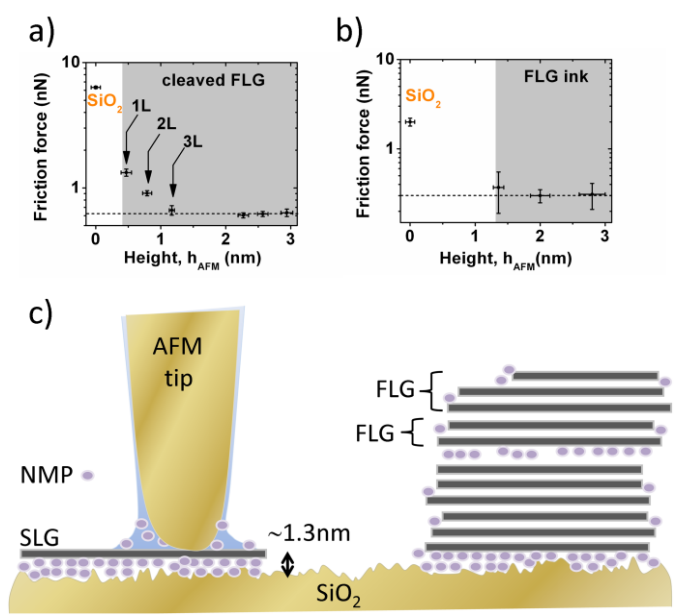

**Fig. 7 (a)** Friction force vs FLG thickness for cleaved flakes on SiO$_2$ ($F_N$ =10 nN). Each data point is an average of several friction data (10-20 points). Clearly, friction increases on going from 3 layers-thick (3L) to 1 layer-thick (1L) flakes. **(b)** Friction vs FLG thickness for LPE flakes ($F_N$=4nN). The minimum height, ~1.3 nm, exceeds that of 1L and 2L-thick cleaved flakes and friction varies weakly in the explored height range. Compared to (a), force values are smaller due to differences in $F_N$ and in the tip radius. **(c)** Sketch of the suggested NMP contamination for printed FLG flakes. NMP intercalation systematically increases the apparent height of printed graphene compared to cleaved graphene, and makes flakes less susceptible to out-of-plane deformations. This explains the phenomenology in (b). Also, the AFM tip interacts with the topmost graphene layers through a capillary water bridge with soluble NMP complexes, which provide a reduced tip-flake adhesion (see also main text).

The friction data reported above demonstrate the achievement of ultralow friction in contact junctions involving FLG flakes prepared *via* LPE. This readily emerges from friction maps (Fig. 3(d) and Fig. S1(b)), and is further established through interpolation of the friction *vs* load characteristics with the MD transition model (Fig. 6(b)). The tight relationship between friction and interfacial corrugation,[81] suggests that LPE and MC produced flakes might have similar corrugations of the tip-surface interfacial potential. This is firstly supported by the evidence of a rather regular, atomic-lattice stick-slip friction on LPE flakes. Secondly, the corrugation amplitude $E_0$~0.2 eV-1.1 eV evaluated from stick-slip force profiles on LPE FLG (normal load $F_N$~10 nN) is in the range of literature data for MC SLG/FLG.[12] Contrary to the case of reduced graphene oxide flakes,[38] FLG flakes produced by LPE are not exposed to oxidation and harsh reducing conditions, hence very small amounts of in-plane topological defects and oxygen-related functionalization (hydroxyl and epoxy groups)[59] are in principle expected. For the FLG-based inks considered in this work, XPS analysis indicates the presence of carbon-oxygen groups below 8%. This is certainly one of the main factors explaining the ultralow friction properties of the LPE FLG flakes.

The key finding of ultralow friction is weakly affected by the specific solvent of the ink, in the limit of a reversible solvent-graphene interaction (physisorption).[94] In fact, we studied variable-load friction measurements on vacuum-annealed patterns printed via FLG dispersions in ODCB,[62,64] and obtained shear strength values close to the estimates of Table 1 (see Appendix, section S7). On the other side, we measured ultralow friction only on thermally annealed samples in vacuum at 350°C for 90min. Thermal treatments at lower temperatures (≤250°C) led to inconclusive results due to surface contamination by solvent molecules (see Appendix, Fig. S8). This is in line with reports using thermal treatments above 300°C to effectively remove organic residues from both graphene[101,102] and other carbon-based surfaces.[94]

Another interesting issue concerns the reduction of the adhesive force and effective interfacial energy γ on printed FLG flakes, compared to the case of MC graphene (see Fig. 6(b) and Table 1). Under ambient conditions, the tip-graphene adhesion is dominated by van der Waals[103] and water capillary forces,[95,100] which are in turn greatly influenced by adsorbed surface contaminants.[100,103,104] In this respect, we believe that trace amounts of NMP (below 4% of nitrogen according to XPS analysis) are playing a crucial role. In fact, NMP can reduce the Hamaker constant hence the van der Waals attraction between tip and sample more effectively than adsorbed water molecules.[105] Also, NMP provides water soluble hetero-complexes (NMP($H_2O$)$_2$),[105] which can lower the capillary pressure in the water bridge formed between tip and graphene, thus reducing the capillary component of the adhesive force.[104] For completeness, in Fig. 7(c) we propose a sketch of the possible NMP contamination for printed SLG/FLG. The vacuum annealing procedure at 350° C for 90 min. should remove most of the solvent molecules from the topmost graphene layers (that assures to routinely observe ultralow friction and atomically resolved stick-slip motion by AFM)[94] but it leaves NMP intercalated between the stacked FLG flakes, and at the flakes-substrate interface. This scenario is consistent with previous experiments on NMP-based graphene deposits,[43] and with studies showing that NMP physisorbs on silica[105] and graphene surfaces.[94] Intercalated NMP is expected to play a twofold role. On one hand, it gives a larger height for the LPE flakes compared to the MC ones. Accordingly, one might tentatively ascribe the minimum height ~1.3 nm of Fig. 7(b) to very thin FLG flakes, *i.e.*, SLG or bilayer graphene, adsorbed onto $SiO_2$, with a molecularly-thin layer of NMP trapped at the buried interface. On the other hand, trapped NMP is likely to reduce the flakes-substrate interfacial roughness, making SLG/FLG flakes less prone to out-of-plane deformations around the AFM tip.[13] This might ultimately explain the weak thickness-dependent friction reported in Fig. 7(b) for the LPE flakes.

**Friction and wear at the printed SLG/FLG edges.** In this section, we focus on the different friction behaviour of the basal plane and surface exposed step edges of FLG flakes. Fig. 8 shows AFM data for a representative aggregate of FLG flakes printed onto $SiO_2$. In the topography of Fig. 8(a) the FLG flakes are easily distinguished. Their random stacking generates a morphology characterized by slightly tilted and smooth terraces separated by well-defined steps, where the surface slope varies abruptly. The friction map in Fig. 8(b) indicates enhanced friction at several FLG flake edges compared to the smooth terraces. Examination of the height profiles and of the related friction loops clarifies that the friction excess originates from the different responses between the trace and retrace scan directions at the step edges (Fig. 8(c),(d) and Appendix Fig. S9). In fact, step up scans show lateral force peaks at the step edges due to the additional resistance experienced by the AFM tip as it moves up the steps, whereas step down scans exhibit smaller, assistive forces helping the tip to move down the steps. As a result, the tip interaction with the surface steps enhances loop hysteresis and friction. This effect qualitatively agrees with experiments and simulations dealing with friction at atomic step edges.[19] Within this framework, it is well known that friction at steps depends on atomic details of the tip-sample junction, as the step geometry, tip apex shape and environmental contaminants adsorbed at the step edges. Hence, we simply observe that friction at the step edges separating neighbouring flakes is increased by a factor ~3 to ~10 compared with the friction measured on the smooth terraces. This fits the typical friction enhancement recorded at the atomic steps of graphite and cleaved SLG/FLG flakes.[18] Remarkably, much higher friction contrast takes place at specific positions of the step edges, as position B of Fig. 8.

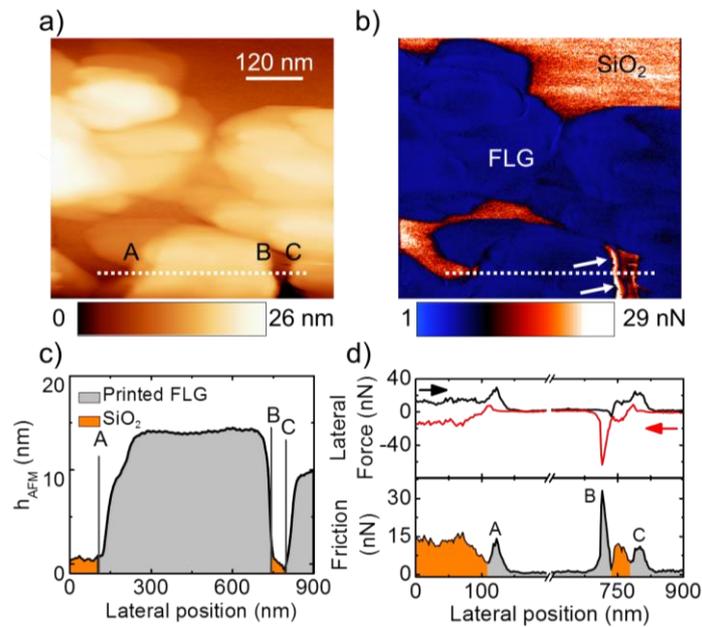

Fig. 8. (a) Topography and (b) friction maps for an aggregate of FLG flakes printed onto $SiO_2$ (normal load $F_N$ =25nN). Friction is enhanced at several surface exposed step edges, as attested by the dark contrast along the edges. White arrows in b) are localized at a high-friction edge. (c) Height profile and (d) friction data acquired respectively along the dash lines in (a) and (b). Peaks at positions A, B and C quantify the friction excess at the step edges. Additionally, the stick-slip instability of force profile at B suggests elastic straining of the edge.

Here, the lateral force peak is ~50 times higher with respect to the force measured on the adjacent smooth terrace (see the corresponding peak in Fig. 8(d)). The occurrence of a stick-slip spike in the related lateral force profile further suggests that the large friction contrast originates from elastic straining of the flake edge[106] along the tip fast scan direction (Appendix, Fig. S9). We note that the strong interaction of the AFM tip with the printed FLG edges is also able to initiate secondary dissipative processes related to structural modifications of the flakes.

In Fig. 9 we illustrate their occurrence during the sequential scanning of individual, FLG flakes. Discontinuities in both topographic (Fig. 9(b)) and lateral force profiles (Fig. 9(c)) mark the onsets of distinct processes, namely tip-induced lateral translation and edge wear, respectively. The appearance of AFM-induced edge deformation and wear for LPE FLG flakes follows similar findings for highly oriented pyrolytic graphite[18] and MC SLG.[106] For MC SLG, edge wear is known to proceed *via* wrinkles formation and partial peeling of graphene from the deposition substrate.[106] Wear starts at sufficiently large values of the normal load $F_N$ (≈40 nN),[106] which are required in order to overcome the adhesion energy of SLG to the substrate. Being far from a comprehensive description of tip-induced wear in printed SLG/FLG flakes, we nonetheless document the occurrence of edge wear events at relatively small $F_N$ values, ≤ 15 nN.

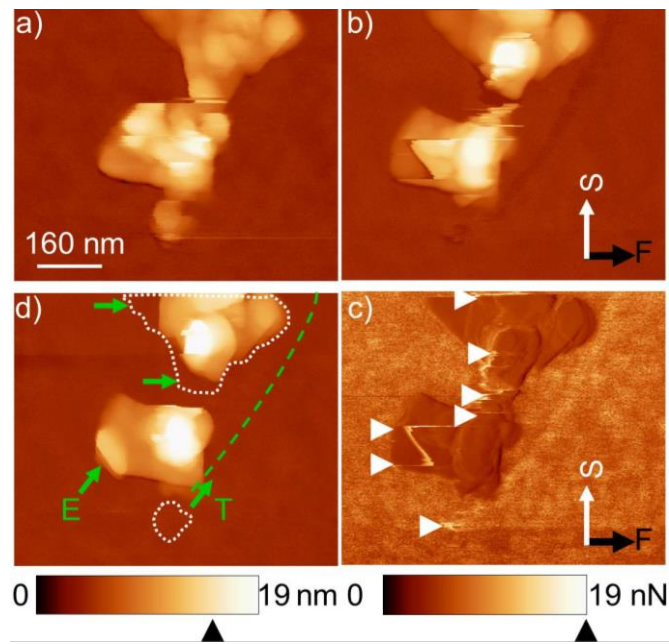

**Fig. 9.** (a), (b) and (d) are AFM topographies acquired sequentially over a region with two aggregates of LPE FLG flakes (substrate HMDS-SiO$_2$, $F_N$ =10nN). (c) is the lateral force map measured simultaneously with (b). The black and white arrows in (b),(c) correspond to the fast F and slow S scan directions. The white triangles in (c) indicate sudden jumps of the lateral force, which mark the onsets of surface structural modifications. The (green) dash line and the (white) dash contours in (d) highlight changes to morphology induced by AFM tip, respectively lateral translation T and edge wear W.

We think that the facile wear in printed SLG/FLG flakes might be a consequence of the scenario depicted in Fig. 7(c), specifically of the reduction of the flake-flake and flake-substrate adhesion energy caused by locally trapped NMP molecules Also, edge defects and/or edge functionalization with adventitious species might enhance wear[18] in printed SLG/FLG flakes. Further insight on these aspects requires the controlled AFM manipulation of SLG/FLG flakes under well-defined load and thickness conditions, which is clearly out of the scopes of the present study.

## Conclusions

In summary, we demonstrate that FLG flakes ink-jet printed onto a solid support are suitable for fundamental friction studies down to the single-flake level. The AFM tip-flake contact shows atomic-scale stick-slip motion and ultralow friction from multi-layers down to few-layers. Contact mechanics modelling gives shear stress values in the 15-29 MPa range against Si tips. The energy landscape experienced by the tip on FLG flakes produced by LPE appears comparable with the MC ones, due to the crystalline quality and minimum amount of in-plane topological defects and oxygen-related species assured by state-of-the-art LPE FLG samples. The latter can be virtually probed on any surface and in combination with other co-deposited nanomaterials. This largely expands the number of systems and the phenomenology that might be investigated through AFM nanomanipulation experiments, in analogy with well-established approaches to nanoparticles friction.[107]

We also demonstrate that the surface exposed step edges represent the main source of friction enhancement and are preferential sites for the emergence of secondary dissipative processes typical of lamellar materials, as edges straining and wear. Moreover, solvent residues probably give rise to the low surface energy and the weak thickness-dependent friction displayed by the ink-jet printed FLG flakes.

High-quality ink-jet printed graphene deposits, with thickness below 100 nm, can be obtained in a few print passes over different substrates.[63,70,108] Surface roughness is slightly higher (by factors ~2-3) than that of the anti-friction, multi-layer thin films typically prepared by drop-casting and dip-coating of LPE graphene, or by wet-transfer of CVD-grown-graphene.[7,20,38,109] In this respect, the capability to deliver ultralow-friction-graphene over technologically relevant substrates, using a scalable production route and the high-throughput, large-area printing technique, might open up new opportunities in the lubrication of micro- and nano-electro-mechanical systems.[110,111] Furthermore, ink-jet printed FLG flakes might reduce stiction, friction, and wear in devices with oscillating, rotating, and sliding contacts.[35,36,112,113]

From a more general perspective, our nanoscale study adds consistency and interest to the preparation of tribofilms from FLG flakes produced by LPE using different coating/printing techniques. It also provides background knowledge for new experiments, "bridging the gap" between the properties of the pristine flakes and the tribology of multi-layer self-assemblies.

## Acknowledgements

This work was supported by the European Union's Horizon 2020 research and innovation program under grant agreement No. 696656—GrapheneCore1 and Regione Liguria PAR-FAS 2007-2013 NEMESI.

# Appendix

## S1. Applicability of the Maugis-Dugdale continuum model to the tip-flake junction

In the present study, the non-linear friction-load dependence of experimental $F_{fric}$ $vs$ $F_N$ curves is traced back to a sublinear increase of the contact area $A$ with the normal load $F_N$, which in turn reflects the reversible elastic deformation of the sphere-on-flat contact geometry. Hence, friction force is assumed to scale linearly with contact area, $F_{fric} = \tau A$, with τ being the interfacial shear strength. This assumption is commonly done to exploit single-asperity contact mechanics in the analysis of atomic force microscopy (AFM) friction force data.[1] Notably, continuum models can effectively interpolate experimental $F_{fric}$ $vs$ $F_N$ curves for different carbon-based surfaces, including graphite, micromechanical cleavage (MC) few-layer graphene (FLG) flakes and multilayer graphene thin films,[2-7] and they can provide reasonable estimates for the contact parameters. The use of the Maugis-Dugdale (MD) model appears therefore particularly convenient. Atomistic simulations of *simplified* model systems however show that continuum contact mechanics does not always capture the rich phenomenology of mechanical interactions taking place in nanosized contacts.[8] Hence the MD continuum model, albeit of widespread use, is intended to provide an oversimplified picture of the actual mechanical interactions between the AFM tip and single layer graphene (SLG)/FLG flakes. In particular the assumption $F_{fric} = \tau A$ can break-down at the nanoscale. Hereafter, we resume for completeness two scenarios that invoke incommensurate contacts, rather than $F_{fric} = \tau A$, to explain the sublinear friction-load dependence observed in AFM experiments. Both might be in principle considered to interpret the experimental friction-load curves, even if the complexity of the studied system - tentatively depicted in the sketch of Fig. 7(c) - does not allow gaining deeper knowledge on this aspect.

The first scenario assumes the tip-graphene junction to behave as a rather incommensurate contact due to the amorphous character of the AFM tip. Atomistic simulations indicate that a bare incommensurate contact (with negligible adhesion) exhibits almost no friction,[8] a condition known as superlubricity or structural lubricity. In such case the friction force increases *linearly* with normal load. The assumption $F_{fric} = \tau A$ no longer holds because the contact is atomically rough, being dominated by the inherent, interfacial atomic roughness.[8,9] van der Waals adhesion is shown to change the $F_{fric}$ $vs$ $F_N$ dependence from linear to sublinear, which is consistent with the MD model and qualitatively agrees with AFM experiments.[9] However, material parameters within the contact region are not properly captured when interpolating such simulations with the MD model.[9] It follows that sublinear friction-load curves have to be explained through a complex interplay of adhesion and atomic roughness effects, rather than through $F_{fric} = \tau A$. Very recent simulations indicate that a comprehensive treatment of the tip-flake junction should also account for the graphene out-of-plane flexibility, as this drives the contact area and the interfacial friction to evolve continuously under applied strains *via* atomic configurational relaxations.[10]

The second scenario assumes that incommensurability at the tip-graphene contact might be caused by a small graphene flake at the tip apex, forming an incommensurate contact with the underlying graphene surface. This picture follows from well-known AFM experiments on graphite,[11] in which peculiar atomically-resolved friction maps were interpreted assuming that a graphite flake was dragged along with the tip and the friction between a graphene flake and the remaining graphite sheet was measured instead of that between the tip and graphite. For such an extended contact, the friction force strongly depends on the orientation of the flake with respect to the substrate. The two contacting surfaces form a superstructure with a Moiré pattern, which gives the relevant friction force. In particular for angles in between commensurate conditions, the potential barriers to sliding are averaged out leading to structural lubricity. Structural lubricity is nonetheless reversible disrupted at certain values of normal load, so that non-linear $F_{fric}$ $vs$ $F_N$ curves can be ascribed to a progressive locking and commensurability transition of the tip flake, as a result of vertical motion of its edge atoms.[12] Again the assumption $F_{fric} = \tau A$ does not strictly hold for a flake mediated tip-graphene contact. As in the present experiment nanosized liquid phase exfoliation (LPE) flakes are studied, one cannot exclude that small flakes might be occasionally captured by the AFM tip apex.

S2. Morphology, adhesion and friction of LPE nanoflakes drop-casted onto SiO$_2$

Figure S1 resumes typical AFM morphology, friction and adhesion data for an individual FLG flake produced by LPE of pristine graphite in N-methyl2pirrolidone (NMP), belonging to the sample prepared by drop-casting onto SiO$_2$ (see sect.2.3 and 2.4 of the main text for details). When imaging individual flakes, F$_N$ is in the range of a few nN, to avoid tip-induced translation and edge wear (see also sect. 3.4. of the main text).

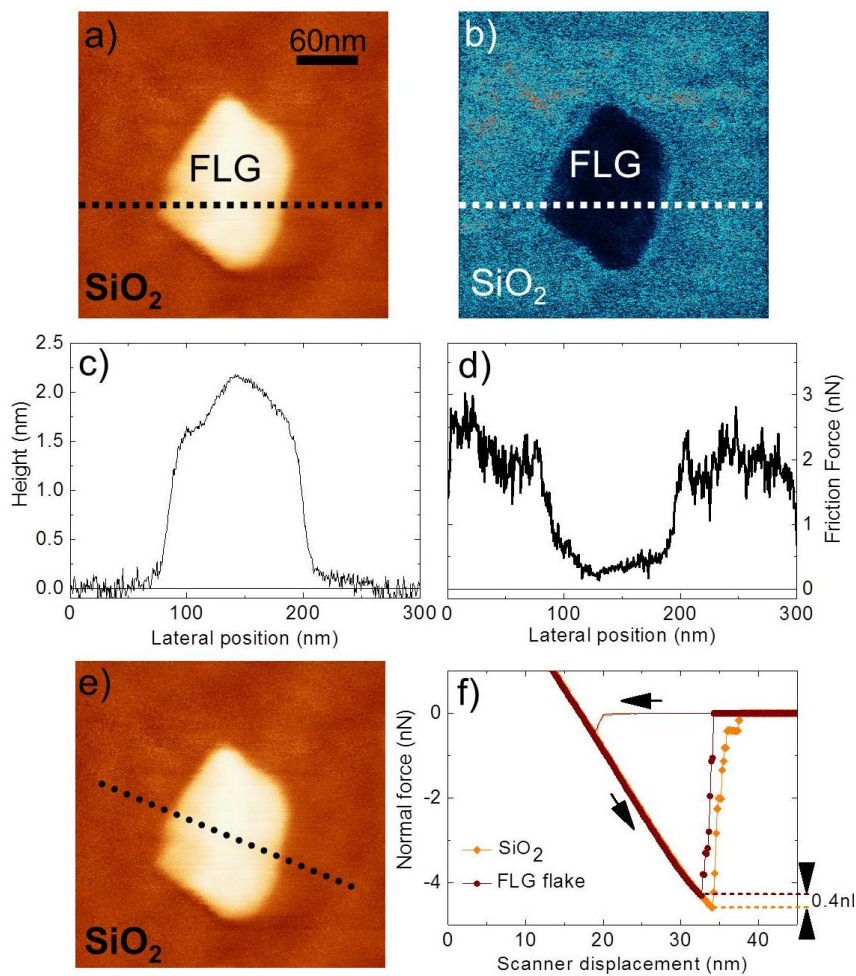

**Figure S1** (a),(b) Topography and friction maps acquired simultaneously on a FLG nanoflake deposited on $SiO_2$ ($F_N$ ~5.1nN). (c),(d) Scan lines corresponding to dash lines in (a) and (b) respectively. They confirm the significant reduction of friction for the ~2nm-thick flake. (e) Individual normal force vs displacement curves are acquired across the flake at the in-plane positions indicated by the black circles. Curves on FLG and $SiO_2$ are separately averaged and reported in (f). The pull-off force on FLG is reduced by ~0.4nN compared to that on $SiO_2$.

## S3. Normal force spectroscopy for FLG printed on HMDS-$SiO_2$

Figure S2 resumes normal force spectroscopy data for FLG printed onto HMDS-$SiO_2$. Briefly, the pull-off (adhesion) force on printed FLG is smaller, or at most comparable, with that measured on the deposition substrate.

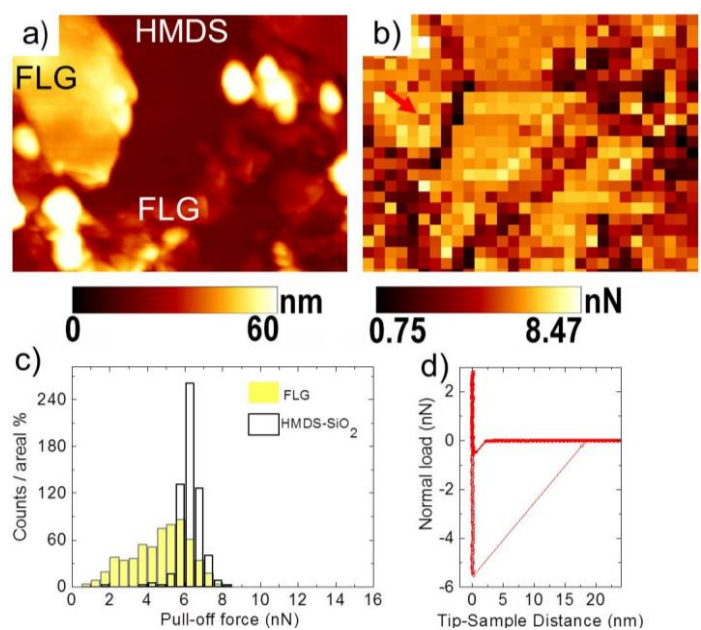

**Figure S2** (a) Topography of FLG printed onto HMDS-SiO$_2$ (size 2.0×1.5μm$^2$). (b) Pull-off force map measured on the aggregate in (a), showing reduced adhesion on FLG with respect to the deposition substrate. (c) Histogram of pull-off forces from the map in (b). (d) Force-distance curve acquired on printed FLG, at the spot indicated by the red arrow in (b). Force profile corresponds to the formation of a mechanically-stiff tip-FLG junction.

S4. Friction response of the HMDS – SiO$_2$ surface

Figure S3 reports representative friction *vs* load curves acquired sequentially over the same surface region of the HMDS-SiO$_2$ substrate. With respect to the characteristics in Figure 6(a), the normally-applied load is here increased from snap-in-contact (~0nN) up to 32nN. A broad friction bump appears above 20nN in scan #1 and shifts towards lower loads in the successive scans (~18nN in #2, ~15nN in #3 and ~12nN in #4). As a result, curves #1 and #4 look substantially different for loads above 12nN. This evolution indicates mechanical instability of the tip-sample junction. This is confirmed by observation of wear debris in the topographies acquired after the scans.

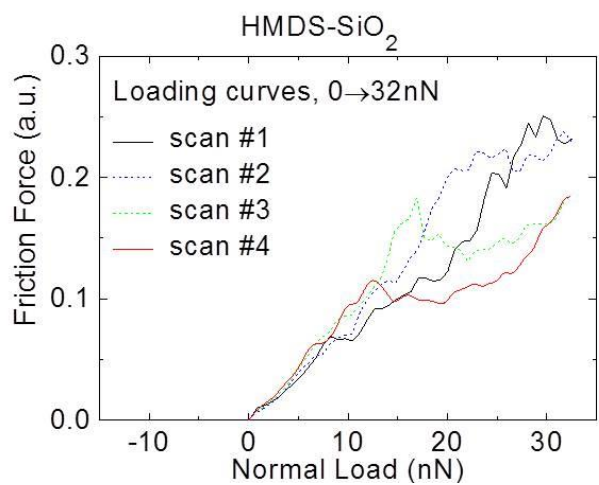

**Figure S3** Evolution of friction *vs* load characteristics (#1 to #4) acquired sequentially over the same surface spot.

S5. Friction response of micromechanically cleaved SLG/FLG

We prepared reference samples of SLG and FLG by MC of graphite on $SiO_2$ substrates. To this purpose we used highly oriented pyrolytic graphite HOPG (ZYB quality by NT-MDT, Russia).

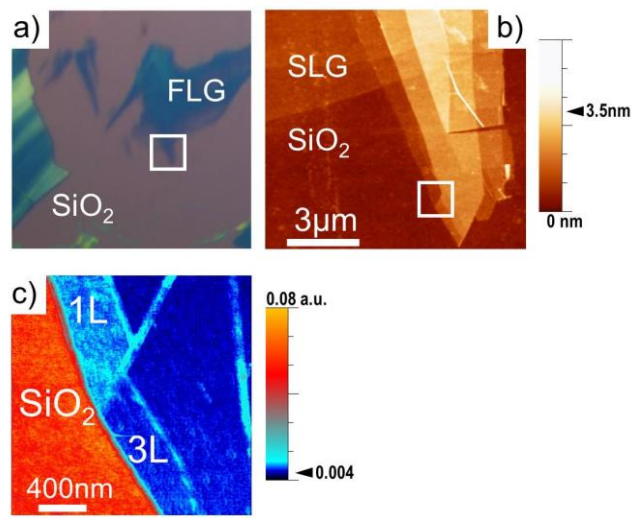

**Figure S4** (a) Optical image of FLG mechanically exfoliated onto $SiO_2$. (b) AFM topography corresponding to the region highlighted in (a), revealing the atomic steps of individual graphene layers. (c) Friction map for the region highlighted in (b). It shows a large friction contrast between $SiO_2$ and graphene, together with thickness-dependent friction for FLG. In fact, friction is clearly higher on the mono-layer (1L) area compared to the tri-layer (3L) region.

After deposition, the thinnest flakes were first identified with optical microscopy, see Figure S4 (a). These flakes have typical lateral dimensions of tens of micrometers. Their thicknesses and friction were systematically measured by AFM. Figure S4 resumes data for a region containing thin and thick flakes.

## S6. Properties of *O*DCB - based FLG ink

The morphology of the flakes produced by LPE of pristine graphite in ortho-dichlorobenzene (*O*DCB) is characterized by TEM (see Figure S5(a) where a representative TEM bright field image is reported). The TEM images allow the estimation of the lateral size of the flakes, with the statistical analysis that shows a main lateral size distribution centred at ~150 nm (Figure S5(a)).

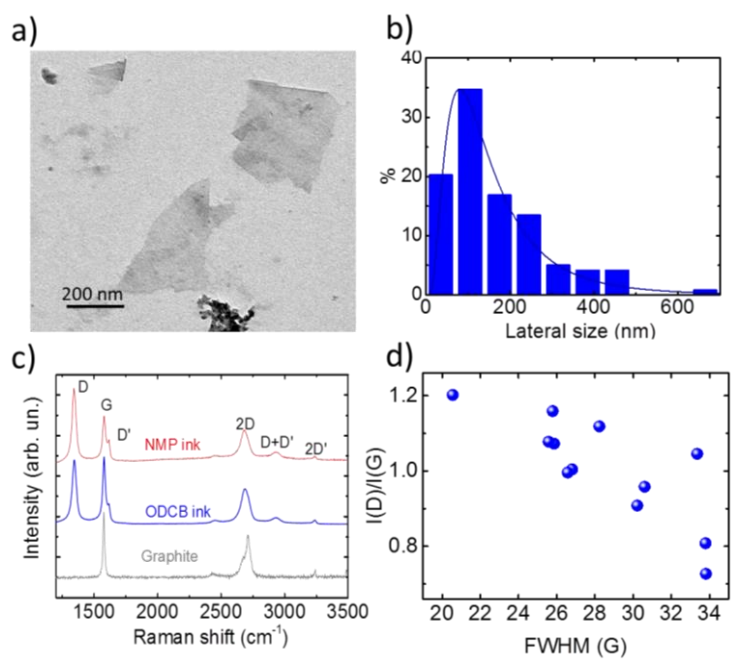

**Figure S5.** (a) Transmission electron microscopy of FLG flakes exfoliated in ODCB and (b) the statistical lateral size distribution. (c) Raman spectrum of the ink in ODCB (blue curve). The Raman spectra of the NMP exfoliated sample (red curve) and the starting graphite (grey curve) are also shown for comparison. (d) The I(G)/I(D) ratio plotted against the FWHM (G) shows no correlation, indicating that the defects are edges rather than structural in-plane.

The Raman spectrum of the as-prepared ink (Figure S5(c)) as well as the spectra of the starting material, *i.e.*, graphite, and the ink in NMP are shown for comparison. The 2D peak gives us information about the number of layers. An estimation of the thickness of the sample can be made by evaluating the $2D_1/2D_2$ ratio, which in our case indicates that our ink is enriched with FLG (Figure S5(c)). The Raman spectrum of the ink produced in *O*DCB also shows significant D and D' intensity, with an average intensity ratio I(D)/I(G)~1, Figure 1 S5(c). This is attributed to the edges of our sub-micrometer flakes, rather than to the presence of basal plane structural defects in the flakes. This observation is supported by the lack of a linear correlation between I(D)/I(G) and FWHM(G). In fact, by combining I(D)/I(G) with FWHM(G) and Disp(G) allows us to discriminate between disorder localized at the edges and disorder in the bulk. In the latter case, a higher I(D)/I(G) would correspond to higher FWHM(G) and Disp(G). Figure S5(d) shows that Disp(G), I(D)/I(G), and FWHM(G) are not correlated, an indication that the major contribution to the D peak comes from the sample edges.

S7. Friction response of *O*DCB-based graphene ink

We probed the friction of FLG printed onto bare SiO$_2$ and HMDS-SiO$_2$, starting from an *O*DCB-based ink. Figure S6 reports a representative, average friction *vs* load characteristic. Interpolation with the MD model gives a shear stress $\tau = 44 MPa$, not far from that measured for the NMP-based graphene inks (see also Table 1 in the main text). A curvature radius R=20nm is used for the calculation.

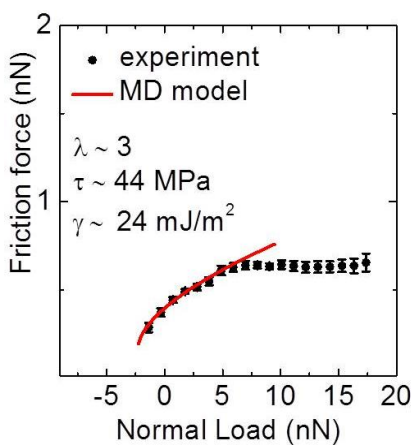

**Figure S6** Average friction *vs* load curve acquired on printed FLG deposited from *O*DCB ink (average over 10 individual curves).

S8. X-ray photoelectron spectroscopy

For detailed analysis about the presence of nitrogen, we carried out the X-ray photoelectron spectroscopy (XPS) N 1s spectra of the deposited samples before and after annealing. The total area of the XPS N 1s core-level spectra of the LPE FLG is normalized by the relative nitrogen amount (N/(N+C)). The nitrogen relative amount is reduced by ca. 35% upon annealing, accompanied by a decomposition of NMP[13] as can be seen by the appearing of a second component at lower binding energy, see Figure S7.

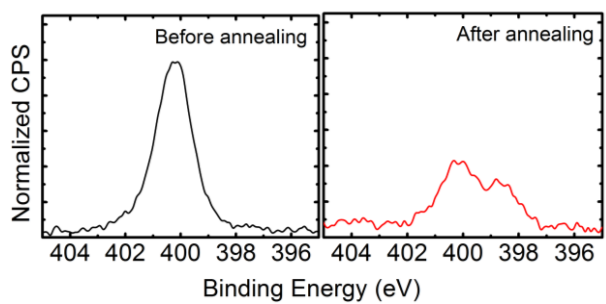

**Figure S7.** The XPS N 1s core-level spectra of the LPE FLG before and after annealing at 350°C in high vacuum.

## S9. Friction of printed FLG vacuum annealed at T≤250°C

Following literature data,[14] we carried out experiments to remove the solvent contamination from printed FLG flakes by 10 up to 60 minutes vacuum-annealing at temperatures T≤250°C. Friction maps demonstrated that such treatment was insufficient to remove solvent molecules from the FLG flakes surface. In fact, under such condition only a fraction of the FLG flakes – as identified from the topographical maps – displayed ultralow friction. The situation is exemplified in Figure S8 for the case of NMP-based ink printed on HMDS-SiO$_2$.

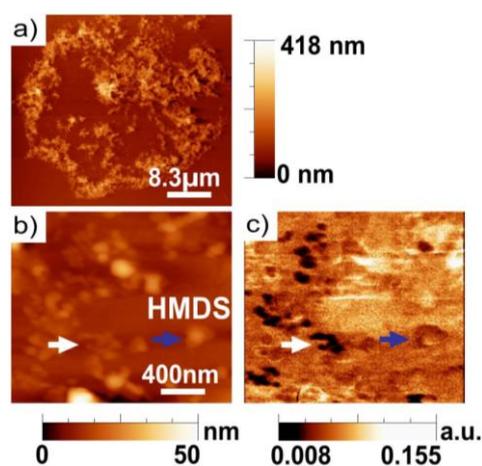

**Figure S8** (a) A FLG microdrop inkjet printed on HMDS-SiO$_2$ using the NMP-based ink and annealed for 10minutes at 170°C. (b) AFM topography and (c) friction maps for a selected region, attesting friction heterogeneity. The white arrow highlights a group of FLG flakes with ultralow friction. The blue arrow on the contrary shows that there are flakes with friction comparable with the deposition substrate. This behaviour reflects massive surface contamination by residual NMP.

## S10. Friction at the printed FLG edges

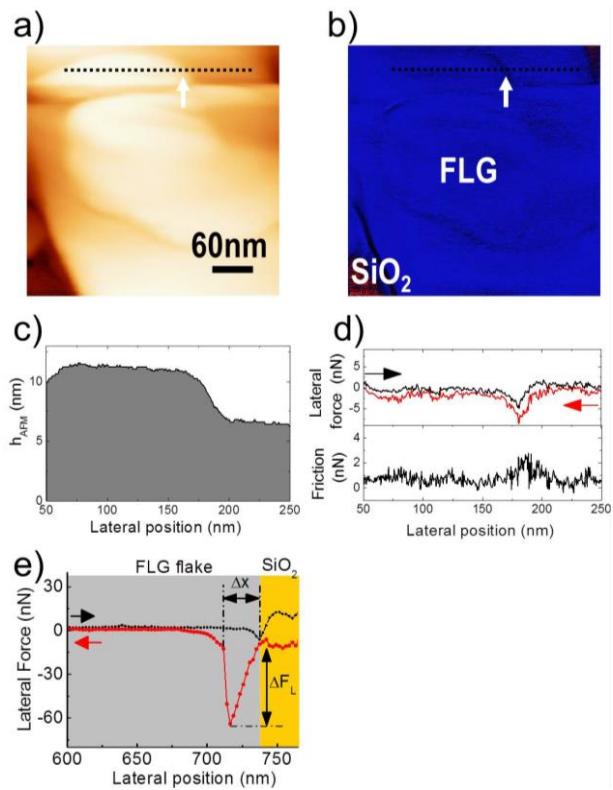

**Figure S9** (a) Topography and (b) lateral force maps for a printed aggregate formed by stacked FLG flakes. A surface exposed step edge is highlighted by the white arrow. (c) Height profile along the dash line in (a) and (b), showing the step edge at the lateral positions between 175nm and 200nm. (d) Lateral force and friction profiles along the dash line reported in (a) and (b), respectively. A friction spike occurs at the position of the step edge. (e) Magnification of the lateral force profile shown in Fig. 8(d) of the main text.

The stick-slip spike in Figure S9(e) has a width $\Delta x \sim 30$nm and amplitude $\Delta F_L \sim 55$nN. The effective spring constant for flexing the FLG edge, estimated from the linear response of the lateral force, is ~1.8N/m. This value is nearly 16 times smaller than the torsional spring constant of the AFM probe $k_{tip} = GJ/[l(h+t/2)2] \approx 29$ N/m, where G = 64GPa is the shear modulus of silicon, J is the torsion constant (approximated as $0.3wt^3$, where w is the 35µm width and t is the 2µm thickness of the cantilever), l is the 350µm length of the cantilever, and h is the 22µm height of the tip apex. Hence, the majority of the deflection $\Delta x$ occurs within the FLG flake when the tip is laterally pressed against its edge at position B.

The effective spring constant of ~1.8N/m overcomes by a factor 6 to 9 that reported for MC SLG (0.2N/m - 0.3N/m). We believe that force increase might reflect, in the present case, elastic straining of more than one graphene layers.

## References appendix